%% file: main.tex
\documentclass[paper]{ieice}
\usepackage[pdftex]{graphicx,xcolor}
\usepackage{my_packages}
\usepackage{my_command_definitions}

\field{}
\title{Efficient and Robust Lock-Free Multi-Word Compare-and-Swap via Contention-Aware Helping}
\authorlist{%
 \authorentry{Motoki Unno}{n}{NU}\MembershipNumber{}
\authorentry{Kento Sugiura}{n}{NU}\MembershipNumber{}
\authorentry{Yoshiharu Ishikawa}{m}{NU}\MembershipNumber{9411330}
}
\affiliate[NU]{Graduate School of Informatics, Nagoya University, Nagoya-shi}
\received{2024}{1}{1}
\revised{2024}{1}{1}

\begin{document}
\maketitle
\begin{summary}
    Efficient concurrent access to shared memory remains a central focus for researchers seeking to enhance data structure performance.
    Lock-based synchronization often limits scalability and introduces liveness issues such as deadlocks.
    In contrast, implementing non-blocking structures with single-word compare-and-swap (CAS) instructions increases algorithmic complexity because of unavoidable intermediate states.
    Multi-word compare-and-swap (MCAS) operations offer a practical primitive for atomically updating multiple discrete memory locations, thereby addressing these challenges.
    However, under high contention, helping mechanisms designed to guarantee lock-freedom may cause excessive cache invalidations and significant performance degradation.
    Furthermore, existing approaches are vulnerable to the ABA problem.
    Current lock-free MCAS algorithms may duplicate the execution of the same operation, leading to inconsistent states in certain edge cases.
    To address these challenges, this paper introduces a new lock-free MCAS algorithm that achieves both efficiency and consistency.
    First, we propose a contention-aware helping mechanism that dynamically regulates the number of concurrent helpers through exponential backoff and embedded entry counters.
    These counters also enable a fast garbage-collection path, significantly reducing memory management overhead.
    Second, we introduce a version embedding approach to suppress the ABA problem during MCAS operations.
    Although version embedding requires several bits per target memory region to store version information, embedded versions allow helpers to avoid duplicated MCAS executions.
    Experimental results show that the proposed method achieves up to three times the throughput of the state-of-the-art lock-free MCAS algorithm.
    Moreover, the results indicate that version embedding is sufficient to prevent the ABA problem in practical scenarios.
\end{summary}

\input{contents.tex}

\bibliographystyle{ieicetr}% bib style
\bibliography{reference}
% \begin{thebibliography}{99}% more than 9 --> 99 / less than 10 --> 9
% \end{thebibliography}

\profile{Motoki Unno}{%
    is a Master's student in the Graduate School of Informatics, Nagoya University.
    He received his Bachelor of Informatics degree from Nagoya University in 2025.
    His research interests include concurrent programming, non-blocking data structures, and parallel computing.
}
\profile{Kento Sugiura}{%
    is an Assistant Professor in the Graduate School of Informatics, Nagoya University.
    He received B.S., M.S., and Ph.D. degrees from Nagoya University in 2013, 2015, and 2018, respectively.
    His research interests include indexing techniques, data stream processing, and uncertain data management.
}
\profile{Yoshiharu Ishikawa}{%
    is a Professor in the Graduate School of Informatics, Nagoya University.
    He received B.E., M.E., and Dr. Eng. degrees from University of Tsukuba in 1989, 1991, and 1995, respectively.
    His research interests include database system technologies such as indexing and query processing,
    spatio-temporal databases and data streams, and the integration of AI and DB technologies.
    He is a member of the Database Society of Japan, IPSJ, IEICE, JSAI, ACM, and the IEEE Computer Society.
}

\end{document}

%% file: contents.tex
\section{Introduction}
\label{sec:intro}

Efficient concurrent access to shared data structures is essential for achieving high performance in in-memory databases~\cite{article:Tu2013} and complex index structures~\cite{article:Leis2013}.
Although many systems employ lock-based synchronization, this approach incurs performance overheads and liveness issues, such as deadlocks, which can constrain overall scalability.
In contrast, algorithms utilizing the \emph{compare-and-swap (CAS)} instruction can eliminate the need for locks and facilitate the construction of non-blocking data structures.
Such algorithms and structures are considered \emph{lock-free} if they guarantee system progress as long as at least one thread remains active~\cite{book:Herlihy2020ch3}.
For example, most lock-based structures possess the deadlock-free property, which requires that all threads remain active to guarantee progress; if a lock holder halts or terminates unexpectedly, other threads may be unable to proceed because of an abandoned lock.
Lock-free algorithms maintain progress even in the presence of inactive threads, thereby providing more robust processing.

Although lock-freedom is a highly desirable property, designing correct lock-free algorithms remains a significant challenge.
The CAS instruction atomically compares a single memory word with an expected value and, only if they match, stores the desired value in the target region.
Additionally, because the CAS instruction returns the current (pre-swapped) value, a thread can determine whether the operation succeeded: if the returned value matches the expected value, the CAS succeeded.
The CAS instruction is a powerful primitive with ``universality'' for synchronization in shared memory, enabling an unbounded number of threads to reach consensus~\cite{article:Herlihy1991}.
Consequently, non-blocking algorithms and data structures using CAS have been the subject of extensive research for many years~\cite{article:Herlihy1990,article:Barnes1993,article:Moir1997}.
However, maintaining consistency across multiple memory locations using only single-word operations significantly complicates algorithm design and implementation.

To address this challenge, prior work has proposed a \emph{multi-word compare-and-swap (MCAS)} operation, which atomically swaps multiple words while guaranteeing disjoint-access-parallel execution~\cite{article:Israeli1994,article:Harris2002}.
Because the CAS instruction operates on a single word, it can introduce unavoidable intermediate states within target data structures.
In contrast, MCAS enables the simultaneous swapping of multiple words, thereby eliminating intermediate states and significantly reducing the implementation complexity of lock-free algorithms.
As a result, several recent works, including BzTree~\cite{article:Arulraj2018} and the concurrent programming library kcas~\cite{article:Karvonen2023}, employ MCAS as a core component.

However, current lock-free MCAS implementations encounter two major challenges.
First, the \emph{helping mechanism} required to ensure lock-freedom can degrade performance.
Lock-free MCAS relies on helper procedures to advance intermediate MCAS states; otherwise, abandoned intermediate states may compromise lock-freedom.
Although this mechanism allows MCAS operations to progress even in the presence of slower or suspended threads, helpers may concurrently modify the same memory regions (i.e., MCAS target words) using CAS.
This concurrent modification leads to mutual CPU cache invalidations, thereby degrading overall MCAS performance.
Second, existing lock-free MCAS algorithms~\cite{article:Harris2002, article:Guerraoui2020} exhibit logical vulnerabilities related to the \emph{ABA problem}~\cite{book:Herlihy2020ch10}.
Although the details of this issue are discussed in \cref{sec:consistency}, briefly, the ABA problem among helpers can result in redundant execution of the same MCAS operations.
While this problem is rare due to the complexity of the required conditions, it can cause inconsistent behavior in specific edge cases.

This paper proposes an efficient and robust lock-free MCAS algorithm to address these challenges.
First, a contention-aware approach is introduced to control the helping mechanism.
Unlike existing methods, the proposed approach does not immediately assist intermediate MCAS operations; instead, it verifies whether assistance is genuinely necessary.
This reduces the number of concurrent helpers, thereby mitigating cache contention while preserving lock-freedom.
Furthermore, \emph{version embedding}~\cite{article:Michael1996} is introduced to practically mitigate the ABA problem.
The fundamental cause of the ABA problem is that helpers cannot distinguish whether an expected value remains unchanged or has been modified and subsequently restored.
Although version embedding does not entirely eliminate the ABA problem, experimental results demonstrate that this approach is sufficient to prevent specific edge cases.

The core contributions of this work are summarized below.

\begin{itemize}
    \item The proposed method introduces a contention-aware helping mechanism to MCAS using an entry counter that provides helpers with the current contention state.
          This mechanism enables helpers to select helping or backing off to achieve efficient MCAS while maintaining lock-freedom.
    \item This paper identifies a logical flaw in existing MCAS algorithms.
          \cref{sec:consistency} details how the CASN algorithm results in a duplicated MCAS operation due to its optimization for memory footprint.
    \item The proposed algorithm introduces version embedding to avoid inconsistent MCAS operations.
          Although this approach requires all MCAS target regions to reserve several bits, embedded versions practically prevent the ABA problem in certain edge cases.
    \item The effectiveness of the proposed approach is demonstrated through exhaustive experiments.
          Experimental results show that the proposed approach achieves up to three times the throughput of the state-of-the-art lock-free MCAS algorithm under high contention.
          This performance reaches that of a deadlock-free method that avoids mutual CPU cache invalidations by sacrificing lock-freedom.
    \item We provide reference implementations, including the proposed method and other comparison methods, as a C++ library~\cite{artifact:mcas}.
\end{itemize}

The remainder of this paper is organized as follows.
\cref{sec:related-work} reviews related work, and \cref{sec:overview} provides an overview of the proposed method.
The details of the proposed method are presented in \cref{sec:helping-control,sec:consistency}: \cref{sec:helping-control} discusses the contention-aware helping mechanism, and \cref{sec:consistency} addresses consistency through version embedding.
\cref{sec:evaluation} evaluates the proposed method through experiments, and \cref{sec:conclusion} concludes the paper.

\section{Related Work}
\label{sec:related-work}

This section reviews existing methods for concurrent multi-word modification.
In addition to established MCAS algorithms, this section discusses alternative approaches for atomically reading and modifying disjoint memory regions.

The following discussion assumes that all MCAS target regions can be ordered to prevent procedural deadlocks.
If any MCAS algorithm attempts to swap words in an inconsistent order, it can cause deadlocks.
For example, if one thread swaps target regions A and B, while another thread swaps them in the reverse order, B and A, a conflict arises.
In this scenario, both threads encounter the other thread's intermediate MCAS and are unable to complete both operations.
Therefore, all MCAS operations must adhere to a specific order, such as the sequence of logical memory addresses.

\subsection{Alternative Approaches to Multi-Word Modification}
Several alternatives exist for multi-word modification, including k-compare-single-swap (k-CSS)~\cite{article:Luchangco2003} and the combination of load-link-extended (LLX) and store-conditional-extended (SCX)~\cite{article:Brown2013}.
These methods demonstrate high efficiency for specific data structures but have limitations in general-purpose use.
Although k-CSS can atomically compare multiple words, it swaps only a single word at a time.
Consequently, k-CSS cannot prevent the emergence of intermediate states in lock-free data structures.
LLX and SCX are extensions of the LL and SC instructions that can avoid the ABA problem.
However, using LLX and SCX requires dedicated metadata within the target data structures.

Transactional memory is another alternative for multi-word modification.
There are two approaches to transactional memory: hardware transactional memory (HTM)~\cite{article:Herlihy1993} and software transactional memory (STM)~\cite{article:Shavit1995}.
HTM is more efficient than STM but relies on specific hardware support.
Although STM does not depend on specific hardware, accessing memory through STM introduces unavoidable overhead.
Furthermore, transactional memory is subject to transaction aborts resulting from size limitations or opaque conflicts.
Several methods, including the multiple-compare-multiple-swap (MCMS) operation~\cite{article:Timnat2015}, have been proposed to leverage transactional memory.
However, fundamental limitations persist.

\subsection{MCAS Algorithms}

There are several algorithms to perform MCAS operations.
These algorithms guarantee different types of progress: wait-freedom~\cite{article:Sundell2011,article:Feldman2015}, lock-freedom~\cite{article:Harris2002,article:Guerraoui2020}, or deadlock-freedom~\cite{article:Sugiura2022}.
However, the objective of this work is to achieve high-performance MCAS processing.
Given the inherent complexity of wait-free MCAS algorithms, this section focuses on lock-free and deadlock-free algorithms.

The CASN algorithm~\cite{article:Harris2002} is the first practical MCAS implementation without specific hardware limitations; thus, most existing methods have followed this approach.
For each operation, CASN prepares a \emph{descriptor} containing all necessary information about the corresponding MCAS.
CASN then embeds the descriptor's address into every MCAS target region as a reservation.
If all embeddings succeed, the MCAS operation also succeeds.
If the MCAS succeeds, CASN updates the descriptor's status to SUCCEEDED and replaces the embedded descriptors with the desired values.
Descriptors also serve as the mechanism for ensuring lock-freedom.
If a thread finds an embedded descriptor, indicating the presence of an intermediate MCAS, it can help complete the MCAS using that descriptor.
Although CASN is applicable across various environments, including persistent memory~\cite{article:Wang2018}, it requires nested descriptor embedding that involves both CASN and restricted-double-compare-single-swap (RDCSS) descriptors.
This requirement results in a non-negligible number of CAS instructions per MCAS operation, leading to significant performance overhead.

The AOPT algorithm reduces excessive CAS instructions in CASN to improve performance~\cite{article:Guerraoui2020}.
CASN replaces embedded descriptors with actual values for each MCAS operation, while AOPT retains descriptors within MCAS target regions.
As described above, each descriptor contains all information about the MCAS, including its current status: UNDECIDED, FAILED, or SUCCEEDED.
When a thread finds an UNDECIDED descriptor, it first helps the pending operation to complete.
Otherwise, it can look up the actual value in the descriptor: the expected value for FAILED and the desired value for SUCCEEDED.
This indirect reference eliminates the need for nested descriptor embedding and write-back CAS instructions.
However, this approach introduces trade-offs in both read performance and memory footprint.
Using indirect references incurs unavoidable overhead for each read operation.
Furthermore, because embedded descriptors may remain throughout the process, additional garbage collection is necessary to prevent memory waste.

Deadlock-free MCAS seeks to improve operational efficiency by sacrificing lock-freedom~\cite{article:Sugiura2022}.
In lock-free MCAS, the overhead of helper mechanisms and garbage collection leads to performance bottlenecks under high contention.
To address this issue, the deadlock-free MCAS omits the helping mechanism and waits for intermediate MCAS operations to complete.
This approach does not guarantee lock-freedom, but it avoids mutual CPU cache invalidations and the need for descriptor garbage collection.
As a result, the deadlock-free MCAS has achieved better performance than existing lock-free MCAS algorithms.
The proposed method aims to reach this performance while achieving lock-freedom.

\section{Method Overview}
\label{sec:overview}

This section describes the data structure and the basic procedures of our MCAS algorithm.
The proposed method builds on the deadlock-free MCAS algorithm~\cite{article:Sugiura2022} and introduces two key mechanisms to achieve both efficiency and lock-freedom: \emph{helper control} that reduces cache invalidations caused by conflicts, and \emph{version embedding} that prevents the ABA problem and ensures lock-freedom.
Since the mechanisms and theoretical backgrounds of these optimizations are discussed in the subsequent sections, this section focuses solely on the basic components of the proposed method.

\subsection{MCAS Descriptor Structure}
\label{sec:data_structure}

Similar to existing methods, the proposed method prepares a descriptor for each MCAS operation.
The descriptor reserves each target-word region by embedding its pointer into the regions, thereby linearizing concurrent MCAS operations.
Because MCAS descriptors and their embeddings play a central role in MCAS algorithms, we first describe their data structures and layouts below.

Each target-word region consists of 64 bits and retains either an \emph{actual value} or a \emph{pointer to an MCAS descriptor}.
To distinguish between actual values and descriptor pointers, we use the most significant bit (MSB) as a \emph{control bit}: 0 indicates a value, and 1 indicates a pointer.
If the control bit is 0, the remaining bits contain actual data along with version information for ABA detection (details are described in \cref{sec:consistency}).
If the control bit is 1, the remaining bits contain the address of the corresponding descriptor with metadata used for the helper procedure (details are described in \cref{sec:helping-control}).

\begin{figure}[b]
    \centering
    \small
    \begin{tabular}{|c|c|c|c|}
        \hline
        \multicolumn{4}{|c|}{\textbf{status}}                                                             \\ \hline
        \multicolumn{4}{|c|}{\textbf{\textit{n}}}                                                         \\ \hline
        \multirow{4}{*}{\textbf{targets[ ]}} & addr 1          & expected 1          & desired 1          \\ \cline{2-4}
                                             & addr 2          & expected 2          & desired 2          \\ \cline{2-4}
                                             & ...             & ...                 & ...                \\ \cline{2-4}
                                             & addr \textit{n} & expected \textit{n} & desired \textit{n} \\ \hline
    \end{tabular}
    \caption{MCAS descriptor structure.}
    \label{fig:my_MCASdescriptor}
\end{figure}

\cref{fig:my_MCASdescriptor} illustrates the data structure of MCAS descriptors in the proposed method.
The \texttt{status} field indicates the progress of an MCAS operation and can take one of three states: UNDECIDED, SUCCEEDED, or FAILED.
The \texttt{n} field indicates the number of target words for each MCAS operation.
The \texttt{targets} field is an array that retains information for each target CAS operation.
Each entry has three fields:
\begin{itemize}
    \item \texttt{addr}: The address of a target-word region,
    \item \texttt{expected}: An expected (i.e., old) value in a target region, and
    \item \texttt{desired}: A desired (i.e., new) value.
\end{itemize}
In descriptors, the \texttt{targets} field is statically allocated to accommodate the maximum number of target words allowed per MCAS operation.
The maximum number can be specified via CMake's compile definitions, with 8 as the default value.
Note that each descriptor is aligned to 64-byte addresses (i.e., cache lines) to prevent false sharing~\cite{book:Herlihy2020appendixB} in our implementation.
Thus, regardless of the maximum number of targets, no descriptor shares a cache line.

\subsection{Procedures of MCAS and Read Operations}

\cref{alg:my_mcas,alg:my_read} present the pseudocode for our MCAS and Read operations, respectively.\footnote{
    In the algorithms, a period indicates a reference to a field within a data structure.
}
Similar to existing methods, we prepare the Read algorithm to load the current value from a target-word region.
Since target-word regions may contain descriptor pointers, accessors could misinterpret these pointers as actual data without the Read algorithm.
Then, updaters perform an MCAS operation with the loaded values to atomically swap multiple words.

It is also worth noting that we expect a single CAS instruction to work as shown in \cref{alg:cas}.
Besides, when embedding descriptors, we use test-and-test-and-set (TATAS) operations~\cite{book:Herlihy2020ch7} instead of simple CAS instructions to avoid unnecessary CPU cache invalidation.
However, for simplicity, we just use CAS to refer to TATAS here.

\subsubsection{MCAS Operation}

The proposed MCAS algorithm consists of two phases: the descriptor-embedding phase on lines 2--8 and the finalization phase on lines 10--14.
Note that \cref{alg:my_mcas} alone does not guarantee consistency when helper procedures are executed concurrently; we explain how to ensure consistency in \cref{sec:consistency}.

\textbf{Phase 1 (Embedding a descriptor pointer)}:
A worker thread, which executes a certain MCAS operation, first embeds the pointer to the corresponding descriptor into every target-word region using CAS instructions.
If every region contains the expected values or its own descriptor address, which may be embedded in advance by helpers, the embedding phase succeeds.
Otherwise, if any target regions have been modified or are involved in other MCAS operations, the thread considers the embedding a failure.
Note that our MCAS algorithm does not call a helper procedure for another descriptor because the target regions will be modified by other MCAS operations, which would cause the ongoing MCAS operation to fail.
Thus, the thread just retries from reading the current values, and the Read algorithm will call a helper procedure if needed.
After embedding, the thread updates the descriptor's status (SUCCEEDED or FAILED) by a CAS instruction, which ensures all the threads (including helpers) can agree on the operation's outcome.

\textbf{Phase 2 (Finalizing MCAS by swapping to values)}:
In this phase, the worker thread replaces the embedded pointers with actual values to finalize its own MCAS operation.
If the descriptor's status is SUCCEEDED, the thread stores the desired (i.e., new) values and completes the MCAS operation.
Otherwise, the thread returns to the expected (i.e., old) values to abort the MCAS operation.
Since helpers may perform these replacements concurrently, using CAS instructions is mandatory to avoid redundant updates and concurrency issues.

\begin{algorithm}[t]
    \small
    \DontPrintSemicolon
    \SetKw{Break}{break}
    \KwIn{$desc$ \tcp*[r]{MCAS descriptor}}
    \KwOut{$succeeded$ \tcp*[r]{True on success}}
    \If{$desc.status = \textnormal{UNDECIDED}$}{
        $status \gets \text{SUCCEEDED}$\;
        \ForEach{$t \in desc.targets$}{
            $word \gets \CAS(t.addr, t.expected, desc)$\;
            \If{$word \neq t.expected \land word \neq desc$}{
                $status \gets \text{FAILED}$\;
                \Break
            }
        }
        $\CAS(desc.status, \text{UNDECIDED}, status)$\;
    }
    $succeeded \gets (desc.status = \text{SUCCEEDED})$\;
    \ForEach{$t \in desc.targets$}{
        \uIf{$succeeded$}{
            $\CAS(t.addr, desc, t.desired)$\;
        }\Else{
            $\CAS(t.addr, desc, t.expected)$\;
        }
    }
    \caption{Proposed lock-free MCAS algorithm.}\label{alg:my_mcas}
\end{algorithm}

\begin{algorithm}[t]
    \small
    \DontPrintSemicolon
    \SetKw{Break}{break}
    \KwIn{$address$ \tcp*[r]{Target location}}
    \KwOut{$v$ \tcp*[r]{Actual value in address}}
    \KwOut{$word$ \tcp*[r]{Current state of address}}
    \While{true}{
        $word \gets$ load a value from $address$\;
        \lIf{$word$ contains an actual value}{
            \Break
        }
        Wait briefly before helping the found MCAS\;
        $word_{+} \gets \text{increment the entry counter of } word$\;
        $word' \gets \CAS(address, word, word_{+})$\;
        \If{$word' = word$}{
            $\MCAS(word)$ \tcp*[r]{Help found MCAS}
        }
    }
    $v \gets$ remove a version counter from $word$\;
    \caption{Read algorithm for MCAS target regions.}\label{alg:my_read}
\end{algorithm}

\begin{algorithm}[t]
    \DontPrintSemicolon{}
    \SetKwRepeat{Do}{do}{while}
    \SetKw{Break}{break}
    \KwIn{$address$ \tcp*[r]{Target location}}
    \KwIn{$expected$ \tcp*[r]{Expected old value}}
    \KwIn{$desired$ \tcp*[r]{Desired new value}}
    \KwOut{$word$ \tcp*[r]{Original value at the target location}}
    \( word \leftarrow \) load a word from \( address \) \;
    \If{\( word = expected \)}{%
        store \( desired \) into \( address \)
    }

    \caption{Expected CAS behavior.}\label{alg:cas}
\end{algorithm}

\subsubsection{Read Operation}

The Read algorithm may call a helper procedure to ensure lock-freedom.
However, we avoid helping as much as possible because frequent (and redundant) helping can worsen the entire performance.
We detail helper controls in \cref{sec:helping-control} and only explain the basic procedure here.
Besides, although the Read algorithm returns the current state of a target region ($word$), we explain its usage in \cref{sec:consistency}.

Our algorithm waits for other threads to complete their MCAS operations, if any exist.
Because MCAS operations finish quickly (at most several microseconds) in most cases, we use a momentary sleep call or give an occasion for rescheduling CPU resources to await a found MCAS.
If the same pointer to the descriptor has remained after waiting, we help to complete it.
This procedure continues until an actual value is found and returned.

\section{Contention-Aware Helping for CPU Efficiency}
\label{sec:helping-control}

Executing the helping procedure to guarantee lock-freedom can cause conflicts over the same MCAS operation, potentially leading to excessive cache invalidation.
Deadlock-free MCAS algorithms do not require any help and simply wait for other MCAS operations to complete, if any are found.
However, lock-free MCAS algorithms require helpers to ensure progressiveness, and the helpers try to modify the same memory regions to complete the same MCAS.
Because this invalidates the CPU cache for each helper, excessive cache synchronization reduces overall performance.

To avoid such cache invalidation, we control the frequency of helping procedure calls using exponential backoff~\cite{book:Herlihy2020ch7}.
In contrast to existing lock-free data structures, such as lock-free queues~\cite{article:Michael1996}, the helping procedure for MCAS operations involves complex steps and multiple CAS instructions.
This means that failing to help results in both excessive cache invalidation and wasted CPU cycles.
Thus, after short spinning with NOP instructions~\cite{book:Herlihy2020ch7}, we make a helper sleep for exponential backoff, as shown in \cref{alg:my_read}.\footnote{
    Because the optimal settings for the number of spin-loops and sleep duration depend on the execution environment, we provide compile-time options to set them.
    By default, helpers loop 10 times for spinning and sleep 10 microseconds as the base of exponential backoff.
}
When the same MCAS descriptor remains at the target region after backoff, the helper calls the helping procedure.

Furthermore, helping can be managed more effectively by introducing \emph{entry counters}.
An entry counter is embedded into each target region along with the corresponding descriptor pointer, as shown in \cref{tab:descriptor_bit_field}, and indicates the number of threads that have joined to help the MCAS operation.
That is, even if multiple threads find the descriptor in the same region, they can detect the existence of preceding helpers after backoff.
Thus, threads do not begin helping if a preceding helper exists, and use the number of helpers as the exponent for exponential backoff.
After backoff, if no new helper arrives, threads try to increment the entry counter using CAS and begin helping if it succeeds.
Note that the target index field indicates its index in the MCAS target regions.
Because other threads have already embedded the descriptor pointer in previous target regions, helpers use this field to begin helping from the next target region.

\begin{figure}[t]
    \centering
    \includegraphics{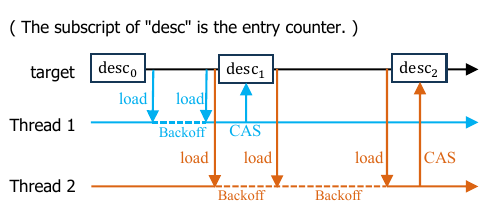}
    \caption{Example of contention-aware helping in execution.}
    \label{fig:backoff_effect}
\end{figure}

\cref{fig:backoff_effect} illustrates how to control the execution of helper procedure calls using an entry counter.
In the example, two threads access the same target region and find the descriptor pointer with the corresponding entry counter.
Since the counter is 0, after backoff, Thread 1 performs a CAS instruction to increment the counter and successfully enters helping.
On the other hand, Thread 2 detects that the counter has been modified and that a new helper has joined to complete this MCAS operation.
Thus, Thread 2 waits again and tries to enter helping after backoff.
In this way, the proposed method limits the number of concurrent helpers, preventing numerous threads from helping the same MCAS operation.

\begin{table}[t]
    \centering
    \small
    \caption{Bit fields for embedding descriptor addresses.}
    \begin{tabular}{ll}
        \toprule
        Bit Range & Content                  \\
        \cmidrule(r){1-1}\cmidrule(l){2-2}
        63        & Control bit (value is 1) \\
        62--50    & Entry counter            \\
        49--47    & Target index             \\
        46--0     & Descriptor address       \\
        \bottomrule
    \end{tabular}
    \label{tab:descriptor_bit_field}
\end{table}

Note that we also use entry counters for efficient memory management.
When there are helpers for a certain MCAS operation, the MCAS descriptor must be freed via any garbage collection (e.g., epoch-based memory reclamation~\cite{thesis:Fraser2004}) to prevent incorrect memory reuse.
However, the owner of the MCAS descriptor can detect the existence of helpers from the entry counters in every target region.
If there is no helper, the owner can store the descriptor in its thread-local storage and safely reuse it for future MCAS operations.
In low-contention (conflicts seldom occur) environments, this approach significantly reduces garbage collection costs and improves CPU and memory efficiency.

\section{Ensuring Consistency Through Version Embedding}
\label{sec:consistency}

Ensuring consistency is a critical challenge in implementing lock-free MCAS algorithms.
In lock-free MCAS algorithms, worker threads may call helper functions and perform the same MCAS operation concurrently.
The existing MCAS algorithms work correctly in most situations, but can lead to inconsistent results due to these helpers.
In this section, we first introduce an example in which the CASN algorithm may redundantly perform the same MCAS operation due to an incorrect combination of its helper function and memory optimization.
Then, we explain how the proposed method ensures consistency using version embedding.

\subsection{Inconsistent MCAS Executions in Existing Methods}

The root cause of inconsistent MCAS executions is the \emph{ABA problem}.
The ABA problem is a well-known issue in programming with CAS instructions: when a thread attempts to perform a CAS instruction with a certain expected value (A), it cannot distinguish whether the current memory state is unchanged (A) or has been modified and restored (ABA).
In MCAS algorithms, the ABA problem arises in descriptor embedding and removal.
For example, when a thread helps an MCAS operation using a found descriptor, it first tries to replace the expected value in the target region with the descriptor pointer.
However, the thread cannot determine whether the target region has been restored to the same value after this MCAS operation completes, or whether it remains truly unchanged.
The thread also cannot determine whether the embedded descriptor is correct if descriptor regions can be reused: another thread may reuse the same descriptor and embed it into the same target region.

The existing algorithms avoid the ABA problem by leveraging the irreversibility of the progress states on descriptors and epoch-based memory reclamation.
The status of a descriptor must be SUCCEEDED or FAILED after completion and must never revert to UNDECIDED.
Even if a helper incorrectly embeds a descriptor, it can recognize the completion of the corresponding MCAS and can stop helping.
Besides, epoch-based memory reclamation prevents other threads from reusing descriptor regions until it ensures that no one is referring to them.
That is, embedded descriptors must not be reused by others, which prevents incorrect descriptor removal from target-word regions.

\begin{figure}[t]
    \centering
    \includegraphics{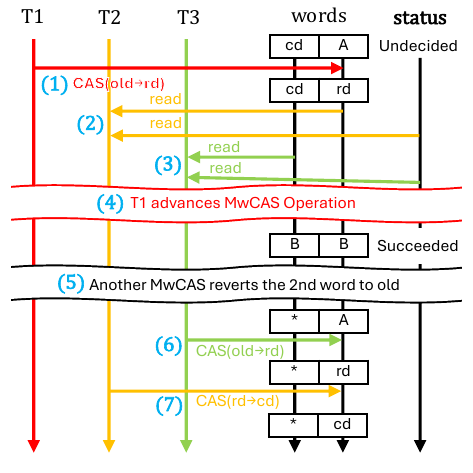}
    \caption{Inconsistency caused by CASN algorithm.}
    \label{fig:CASNbug}
\end{figure}

However, the CASN algorithm may cause the ABA problem due to its memory optimization.\footnote{
    The AOPT algorithm has a similar issue, but we omit its explanation here due to the page limit.
}
We introduce an example in \cref{fig:CASNbug} to show redundant MCAS executions in the CASN algorithm.
In the example, three worker threads perform the same 2wCAS operation.
Thread T1 is the owner of this 2wCAS and embeds the descriptors into the target regions (Step 1).
Note that the CASN algorithm uses two types of descriptors to ensure consistency: CASN (\texttt{cd}) and RDCSS (\texttt{rd}).
Then, T2 finds the RDCSS descriptor in the second target region and confirms that the 2wCAS's status is UNDECIDED (Step 2).
T3 similarly finds the incomplete CASN descriptor in the first target region (Step 3).
Here, suppose that T2 and T3 begin helping but then fall asleep for any reason (e.g., CPU overuse).
T1 embeds the CASN descriptor into the second region and continues until completion during this sleep (Step 4).
After the 2wCAS's completion, other threads can modify the second region and restore the expected (\texttt{A}) value (Step 5).
T2 and T3 may awaken here, but they cannot recognize the completion of the 2wCAS operation because they have already confirmed its status in Steps 2 and 3.
Thus, T3 continues helping by swapping the expected value with the RDCSS descriptor (Step 6), and T2 also swaps the RDCSS and CASN descriptors (Step 7).
As a result, the CASN descriptor is embedded in the second region twice, leading to redundant execution of the 2wCAS operation.

The CASN algorithm can avoid the ABA problem by excluding its memory optimization, but this solution will worsen execution performance and memory efficiency.
In the CASN algorithm, CASN and RDCSS descriptors share the same memory region to use memory efficiently: a CASN descriptor holds RDCSS descriptors for each target in its region.
However, this sharing makes it impossible to determine whether the embedded RDCSS descriptors are correct or are redundantly restored, as they share the same addresses.
If every helper prepares different (i.e., dynamically allocated) RDCSS descriptors, the first and second embedded descriptors into the same region can have different addresses.
This leads to CAS failures in Step 7 in \cref{fig:CASNbug}, but excessive memory allocation and reclamation will decrease overall MCAS performance.

\subsection{Avoiding the ABA Problem via Version Management}

As described above, it is challenging to strictly avoid the ABA problem while achieving high-performance MCAS.
Although the ABA problem can be solved by using massive memory or by sacrificing MCAS performance, we propose a more practical solution by relaxing the restriction.
Inconsistent MCAS executions almost never occur because they require extremely stringent conditions, as shown in \cref{fig:CASNbug}.
However, the existing methods cannot provide information on their safety, so users cannot fully rely on the execution results.
Thus, our approach aims to provide a mechanism for controlling safety rather than ensuring strict ABA avoidance.

In the proposed method, we avoid the ABA problem by embedding a version counter~\cite{book:Herlihy2020ch10} into each target region.
As shown in Step 5 of \cref{fig:CASNbug}, the ABA problem occurs when any thread reverts a target region to its corresponding expected value.
Helper threads cannot determine whether a target region is unchanged without additional information, such as the unique addresses of dynamically allocated RDCSS descriptors in CASN.
Thus, the proposed method maintains the version for each target region as additional information, as shown in \cref{tab:value_bit_field}.
Version counters are incremented when the corresponding MCAS operations successfully update their target regions.
These version changes allow helpers to recognize the corresponding MCAS completions, thereby avoiding the ABA problem.
Note that version counters are only combined with actual values.
Since we can avoid the ABA problem in descriptor removal using descriptor status and epoch-based memory reclamation, embedded descriptor pointers do not have versions and only maintain their metadata, as shown in \cref{tab:descriptor_bit_field}.

Our approach may also lead to inconsistency if the version bit fields overflow and wrap around, but the likelihood can be controlled by changing the number of version bits.
In our implementation, users can specify the number of version bits at compile time.
We use 15 bits as the default setting, as shown in \cref{tab:value_bit_field}, because common OSes provide 48 bits of user-space memory.
We evaluate this setting in \cref{sec:evaluation} and show that 15 bits are sufficient to prevent the ABA problem while supporting the implementation of lock-free data structures.

\begin{table}[b]
    \centering
    \small
    \caption{Bit fields for embedding version counters alongside actual values.}
    \begin{tabular}{ll}
        \toprule
        Bit Range & Content                  \\
        \cmidrule(r){1-1}\cmidrule(l){2-2}
        63        & Control bit (value is 0) \\
        62--48    & Version                  \\
        47--0     & Actual value             \\
        \bottomrule
    \end{tabular}
    \label{tab:value_bit_field}
\end{table}

Below, we explain how versions are managed during the proposed MCAS algorithm.

\textbf{Descriptor Preparation}:
Our Read algorithm fetches both the actual value and the embedded version counter from a target region.
While existing methods do not embed metadata and directly use actual values as expected ones, our approach requires entire words---including version counters---for future MCAS operations.
Thus, our Read algorithm returns the actual value and the entire word, as shown in \cref{alg:my_read}, to compute a desired value from the actual value and use the entire word as the expected state.

\textbf{Descriptor Embedding}:
When each thread embeds a descriptor into the target regions, it uses the expected word states---including version counters---for CAS instructions.
Thus, even if a certain target region is modified and reverted to its expected value, the corresponding version counter will detect it and safely ensure consistency.
Note that if a thread finds other MCAS descriptors in its target regions, our MCAS algorithm immediately fails and does not help those operations, as shown in \cref{alg:my_mcas}.
Although the found MCAS operations may fail and keep the target regions unchanged, in most cases, they will modify the actual values or the version counters.
Therefore, to eliminate wasteful processing, our MCAS algorithm fails early and prompts the thread to retry from reading the target regions.

\textbf{Descriptor Removal}:
When an MCAS operation succeeds, the version counters of the corresponding target regions are incremented by 1 during descriptor removal.
Otherwise, the version counters remain unchanged during MCAS aborts.
The ABA problem occurs only when the target regions are swapped and reverted; it first requires that MCAS operations succeed.
Thus, it is sufficient to update the version counters upon MCAS success.
Note that the version counters are embedded within the expected word states in a descriptor, allowing helper threads to reference and increment them as well.

\section{Evaluation}
\label{sec:evaluation}
This section evaluates the performance and robustness of the proposed method.
First, the experimental setup and methodology are described.
Next, the proposed method is compared to existing MCAS algorithms in terms of throughput and latency.
Furthermore, the effectiveness of the consistency guarantee provided by the version counter embedding is evaluated.

\subsection{Experimental Setup}
The server specifications used for the experiments are presented in \cref{tab:environment}.

\begin{table}[b]
    \caption{Experimental environment.}
    \label{tab:environment}
    \centering
    \begin{tabular}{ll}
        \toprule
        Item     & Value                                              \\
        \cmidrule(r){1-1} \cmidrule(l){2-2}
        CPU      & Intel(R) Xeon(R) Gold 6258R (two sockets)          \\
        RAM      & DIMM DDR4 (Registered) 2933 MHz (16GB $\times$ 12) \\
        OS       & Ubuntu 22.04.5 LTS                                 \\
        Compiler & GNU C++ ver.\ 11.4.0                               \\
        \bottomrule
    \end{tabular}
\end{table}

All related artifacts, including the proposed method and comparison methods, were implemented in C++~\cite{artifact:mcas}.
The proposed method (denoted as Ours) is compared against three implementations: CASN, AOPT, and the deadlock-free MCAS (denoted as DLF).

Our benchmark program repeatedly executes MCAS operations until a 10-second timeout is reached.
MCAS target regions consist of an array of one million words, initially set to zero.
The logical address of each word is aligned with the cache line size (64 bytes) to prevent false sharing~\cite{book:Herlihy2020appendixB}.
Each MCAS operation randomly selects a specified number of words and atomically increments their values by one.
The selection probability for each word follows a Zipf distribution with parameter $\alpha$, defined as:
\begin{equation}
    \label{eq:zipf}
    f(k; \alpha, |W|) = \frac{1 / k^{\alpha}}{\sum_{n = 1}^{|W|} 1 / n^{\alpha}},
\end{equation}
where $|W|$ is the size of the target regions ($10^6$).
When $\alpha=0$, the selection probability is uniform, which is referred to as low contention.
When $\alpha=1$, the access probability to specific words (the $k$-th region) is concentrated in proportion to $1/k$, which is referred to as high contention.

The experiments were conducted five times per condition, and the average values for throughput and latency were calculated.
Error bars represent the minimum and maximum values of the five measurements, respectively.
Unless otherwise specified, the 99th percentile latency is used as the latency metric to compare the effect of tail latency.

Hereafter, an MCAS operation targeting $n$ words is referred to as CAS$n$; for example, a 2-word CAS is CAS2.
Although experiments were performed with different word counts between CAS1 and CAS8, the results exhibit similar behavior and are thus omitted.
Therefore, only results with CAS2 are presented below.

\subsection{Performance Scaling with Thread Count}

\begin{figure}[b]
    \centering
    \begin{minipage}{0.495\linewidth}
        \centering
        \includegraphics[width=\linewidth]{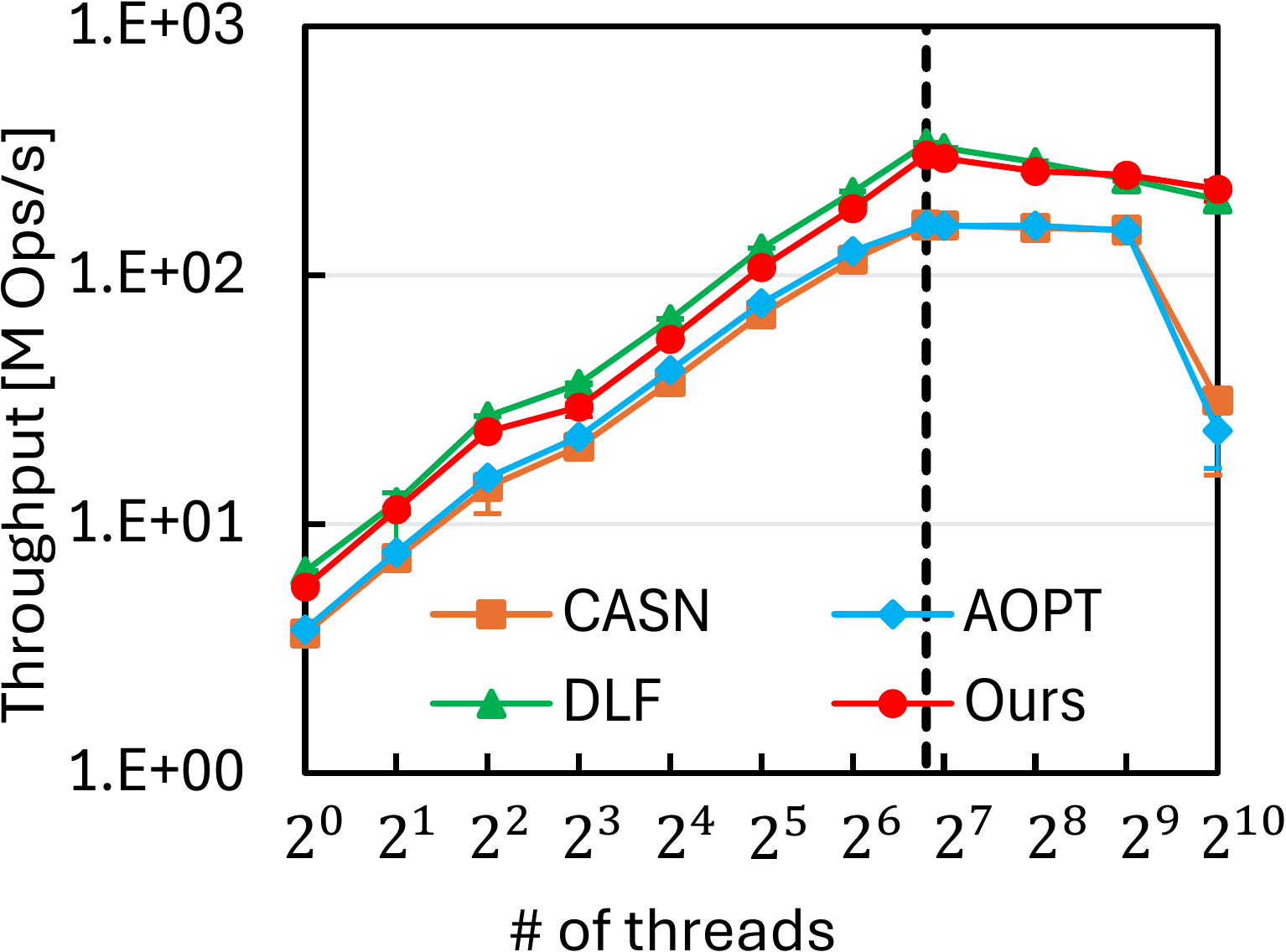}
        \subcaption{Throughput}
        \label{fig:threadsnum_throughput_low}
    \end{minipage}
    \hfill
    \begin{minipage}{0.495\linewidth}
        \centering
        \includegraphics[width=\linewidth]{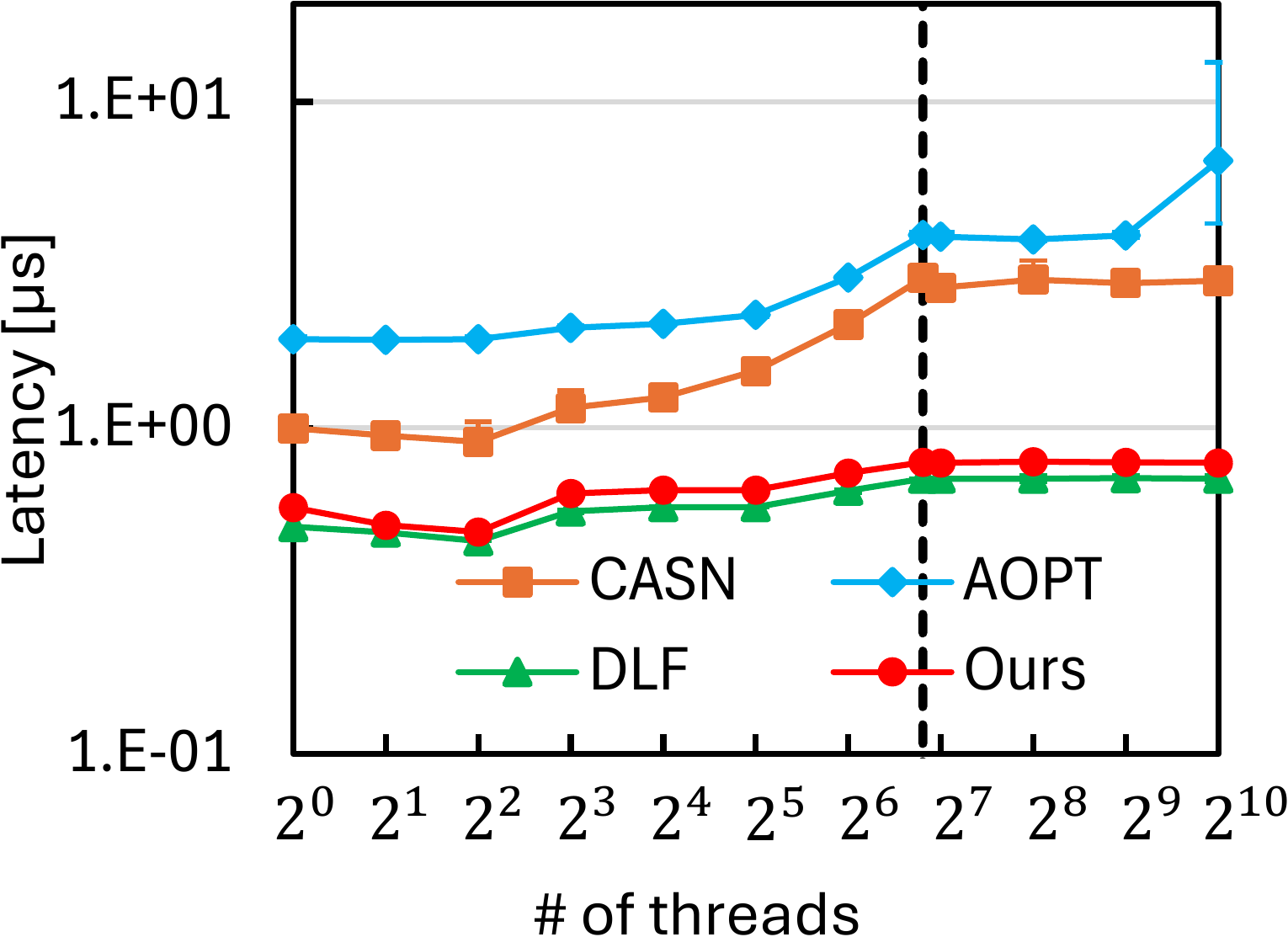}
        \subcaption{Latency}
        \label{fig:threadsnum_latency_low}
    \end{minipage}
    \caption{Performance comparison of CAS2 across different thread counts under low contention ($\alpha = 0$).}
    \label{fig:threadsnum_low_combined}
\end{figure}

\begin{figure}[b]
    \centering
    \begin{minipage}{0.495\linewidth}
        \centering
        \includegraphics[width=\linewidth]{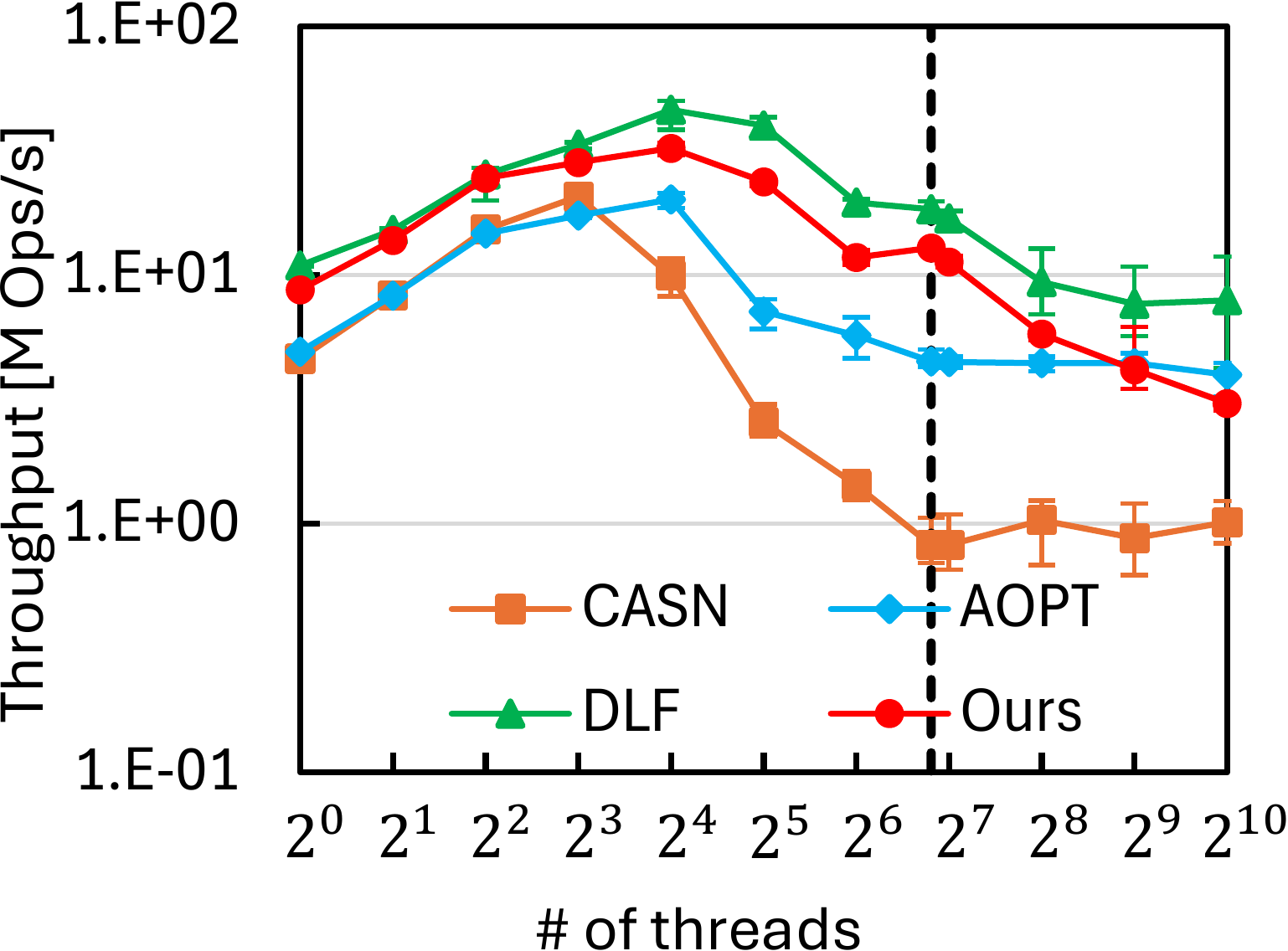}
        \subcaption{Throughput}
        \label{fig:threadsnum_throughput_high}
    \end{minipage}
    \hfill
    \begin{minipage}{0.495\linewidth}
        \centering
        \includegraphics[width=\linewidth]{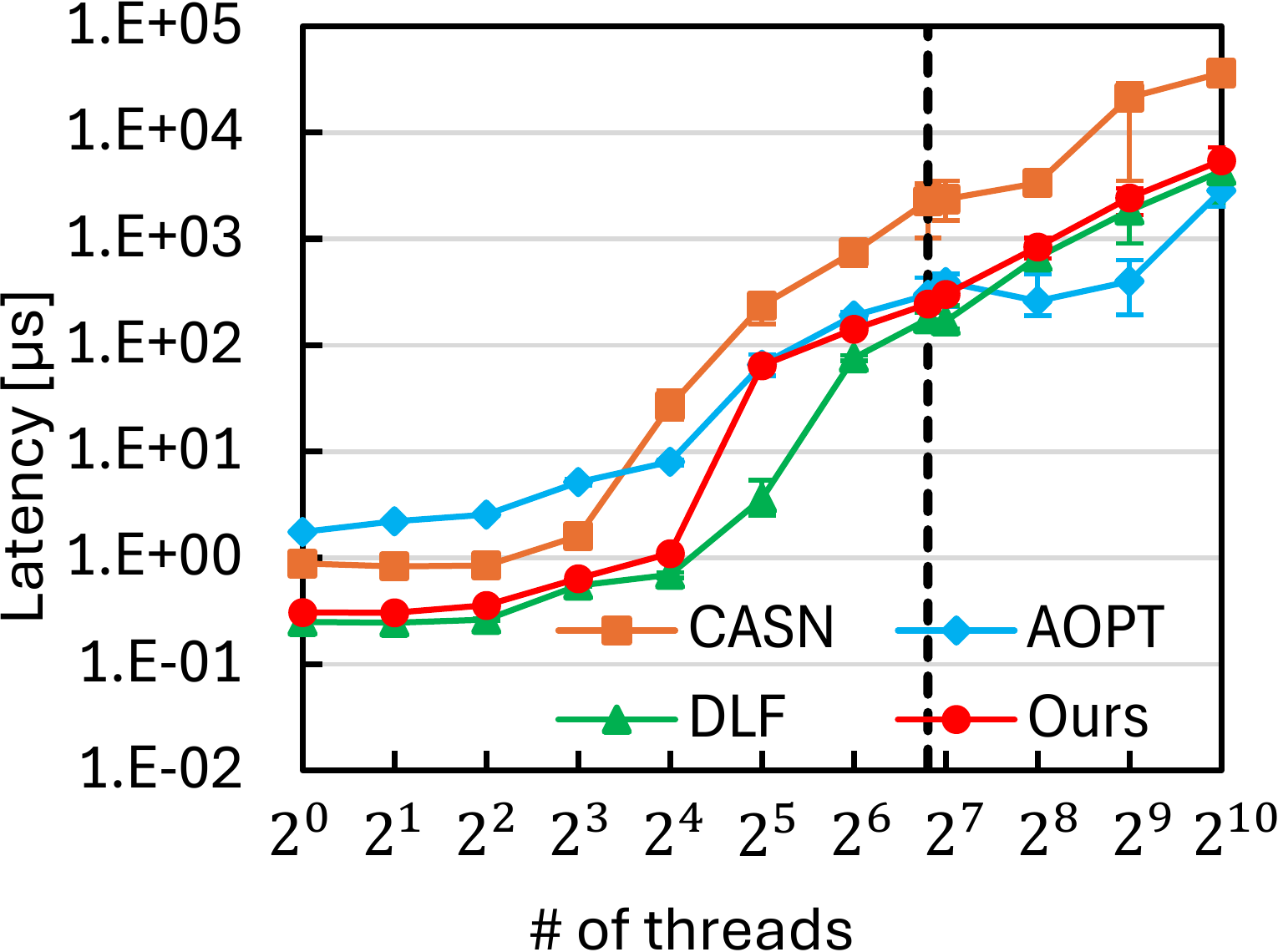}
        \subcaption{Latency}
        \label{fig:threadsnum_latency_high}
    \end{minipage}
    \caption{Performance comparison of CAS2 across different thread counts under high contention ($\alpha = 1.0$).}
    \label{fig:threadsnum_high_combined}
\end{figure}

\Cref{fig:threadsnum_low_combined} shows the throughput and latency of CAS2 under low contention as the number of threads increases.
Vertical dashed lines indicate the boundaries between conditions with and without oversubscription.
The proposed method consistently achieves higher throughput than the existing lock-free MCAS algorithms, reaching approximately twice the throughput at 112 threads.
This performance is close to that of DLF.
Furthermore, while CASN and AOPT's throughput decreases with oversubscription (1,024 threads) due to their complexity and memory management, the proposed method maintains its efficiency.
Regarding latency, the proposed method is also consistently lower; at 112 threads, its latency is about one-fourth that of CASN.
While CASN and AOPT show increased latency as the thread count grows, the proposed method remains nearly constant and close to DLF.

\Cref{fig:threadsnum_high_combined} shows the results for CAS2 under high contention.
Without oversubscription, the proposed method reaches the performance of DLF.
When using 112 threads, the proposed method achieves 12- and 3-times higher throughput than CASN and AOPT, respectively.
Additionally, the proposed method's latency is approximately one-tenth that of CASN and slightly lower than that of AOPT.
These results suggest that the contention-aware helping mechanism successfully mitigates performance degradation during contention.
With oversubscription, the proposed method's performance decreases because it assists with backoff, which can delay MCAS completions.
Even if helpers complete a given MCAS, its owner may continue assisting other MCAS operations and may sleep due to conflict.
However, the proposed method maintains comparable or higher throughput than the existing lock-free MCAS algorithms under oversubscription.

\subsection{Robustness against Contention Skew}

\begin{figure}[b]
    \centering
    \begin{minipage}{0.49\linewidth}
        \centering
        \includegraphics[width=\linewidth]{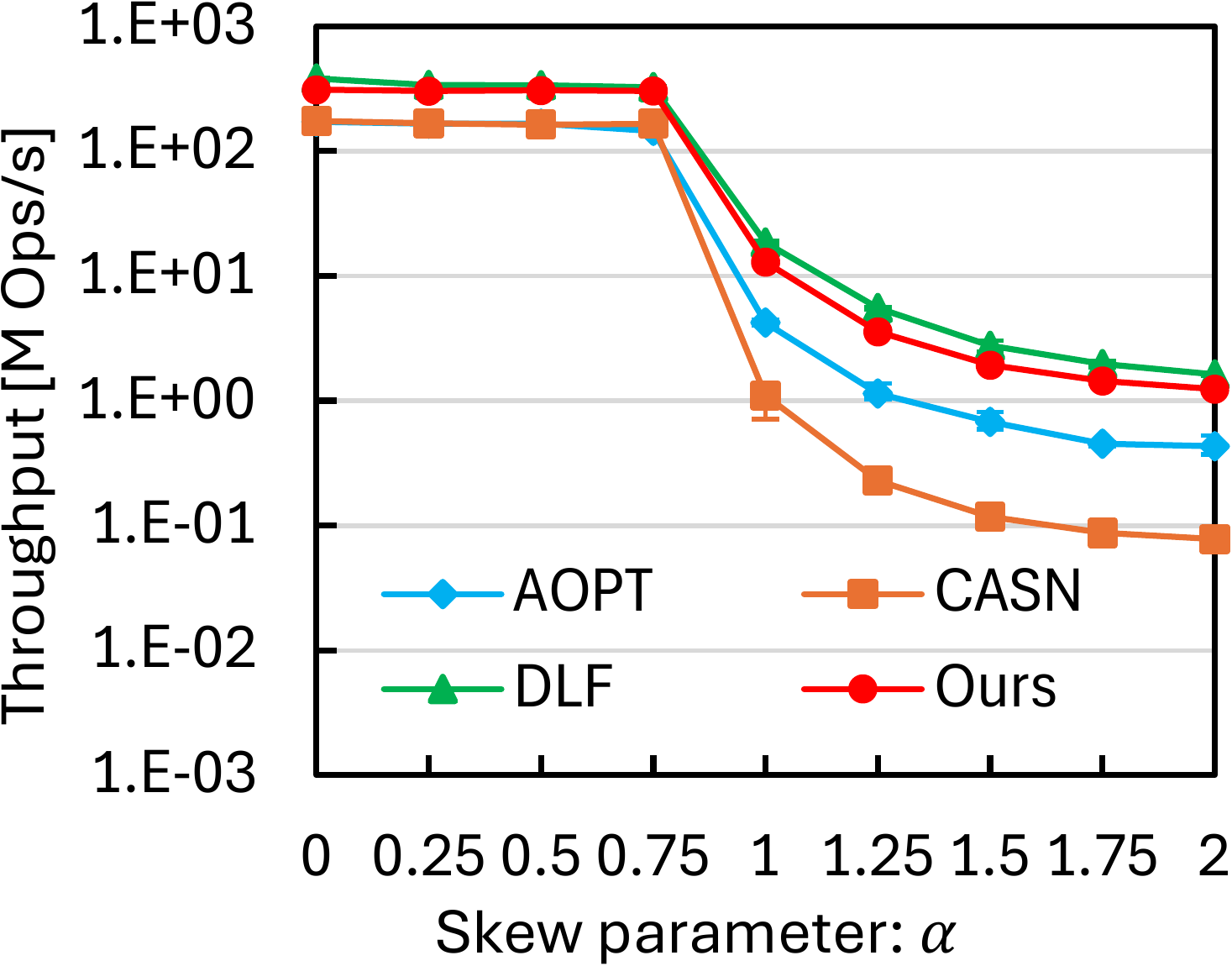}
        \subcaption{Throughput}
        \label{fig:skew_throughput}
    \end{minipage}
    \hfill
    \begin{minipage}{0.49\linewidth}
        \centering
        \includegraphics[width=\linewidth]{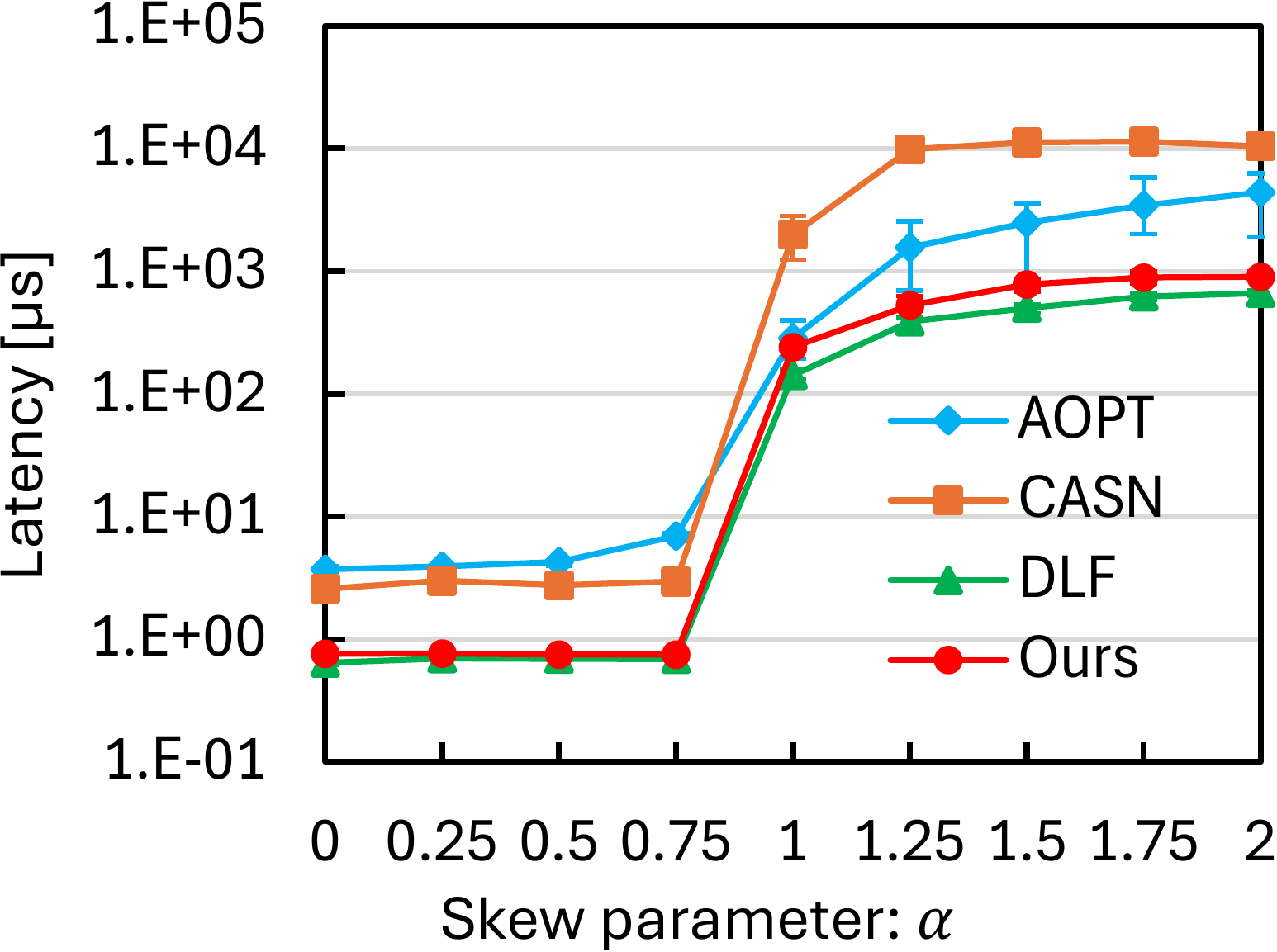}
        \subcaption{Latency}
        \label{fig:skew_latency}
    \end{minipage}
    \caption{Performance comparison of CAS2 across different skew parameters without oversubscription (112 threads).}
    \label{fig:skew_combined_112}
\end{figure}

\begin{figure}[b]
    \centering
    \begin{minipage}{0.49\linewidth}
        \centering
        \includegraphics[width=\linewidth]{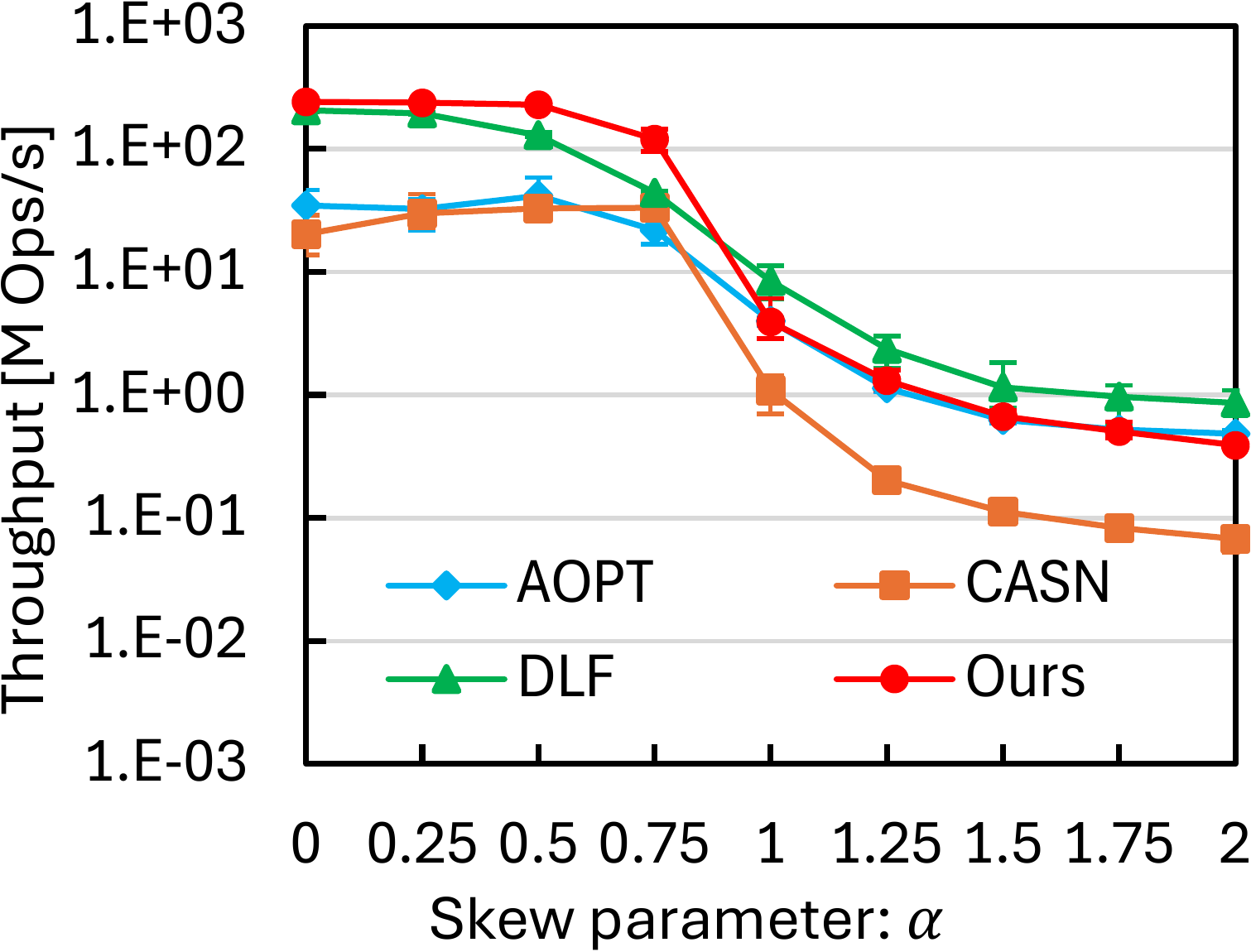}
        \subcaption{Throughput}
        \label{fig:skew_throughput_1024}
    \end{minipage}
    \hfill
    \begin{minipage}{0.49\linewidth}
        \centering
        \includegraphics[width=\linewidth]{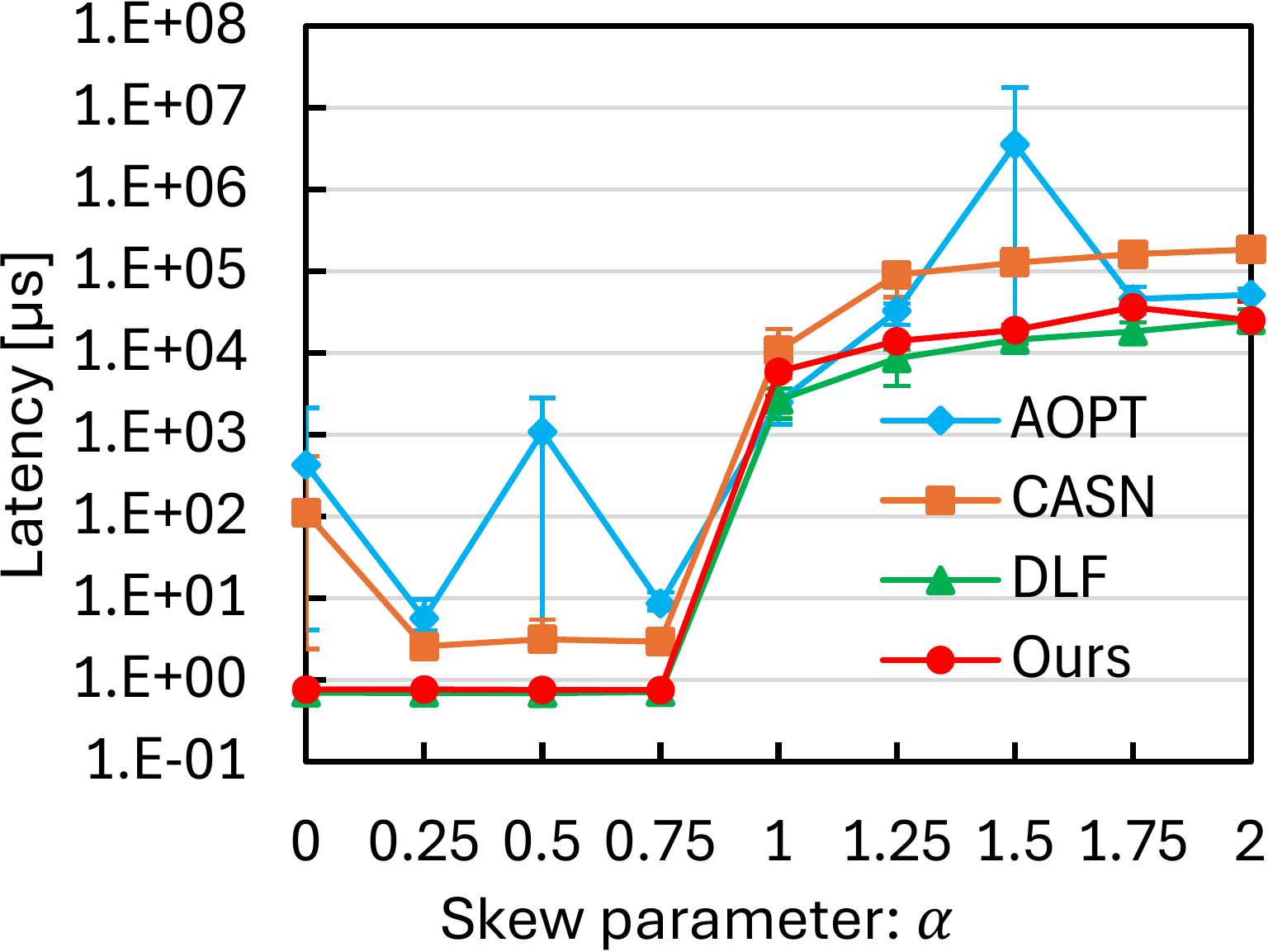}
        \subcaption{Latency}
        \label{fig:skew_latency_1024}
    \end{minipage}
    \caption{Performance comparison of CAS2 across different skew parameters with oversubscription (1,024 threads).}
    \label{fig:skew_combined_1024}
\end{figure}

\Cref{fig:skew_combined_112} shows experimental results for CAS2 with 112 threads (i.e., without oversubscription) over different skew parameters.
Similar to existing methods, the proposed method performs well at low contention but deteriorates rapidly as contention increases.
However, the proposed method consistently outperforms the existing lock-free MCAS algorithms and achieves performance close to that of DLF.
This indicates that the contention-aware helping mechanism effectively alleviates contention and performance drops.

\Cref{fig:skew_combined_1024} shows experimental results for CAS2 with 1,024 threads (i.e., with oversubscription).
With some skew parameters, such as around 0.75, the proposed method achieved slightly higher throughput than DLF.
Although DLF avoids excessive cache invalidations by eliminating the helping mechanism, under oversubscription, it allows incomplete MCAS descriptors to remain indefinitely and obstruct other operations.
In contrast, the proposed method mitigates excessive cache invalidations through backoff while allowing the helping mechanism to remove those obstructing descriptors, thereby improving performance.
Although DLF is superior to the proposed method under higher-contention settings, the performance difference is slight.
Furthermore, the latency results show the robustness of the proposed method.
CASN and AOPT exhibit unstable behavior due to their memory management, such as descriptor allocation and garbage collection.
In contrast, the proposed method avoids unnecessary allocation and reclamation, resulting in robust tail latency.

\subsection{Empirical Safety Analysis of Version Embedding}
Finally, the effectiveness of the consistency guarantee provided by version counters is evaluated.
While version counters mitigate the ABA problem, inconsistencies may arise when the counter wraps around.
Thus, experiments were conducted to identify environments where such inconsistencies might occur.

Two metrics were measured in each environment:
\begin{itemize}
    \item \textbf{Helping Latency ($\mathit{helping\_latency}$)}: The average of the maximum times required for the helping process.
    \item \textbf{Wraparound Interval ($\mathit{wraparound\_interval}$)}: The average time required for the version counter to wrap around.
\end{itemize}
Each metric is not displayed if there is no helping process or wraparound.
If $\mathit{helping\_latency} > \mathit{wraparound\_interval}$, an inconsistency can potentially occur because a helper may encounter a wraparound version.
Otherwise, since no inconsistency can occur, the version embedding can ensure consistency in practice.

Fifteen bits were used for version counters in this experiment, as a pessimistic estimate of the number of bits required to guarantee consistency.
Although the optimal number of version bits depends on the application, this setting will demonstrate that 15 bits are sufficient to ensure consistency even under excessive MCAS operations.
Note that while every thread performs MCAS operations continuously in the benchmark, real applications use them only when necessary.
This results in fewer MCAS operations, preventing version counters from wrapping around in most use cases.

\begin{figure}[t]
    \centering
    \begin{minipage}{0.495\linewidth}
        \centering
        \includegraphics[width=\linewidth]{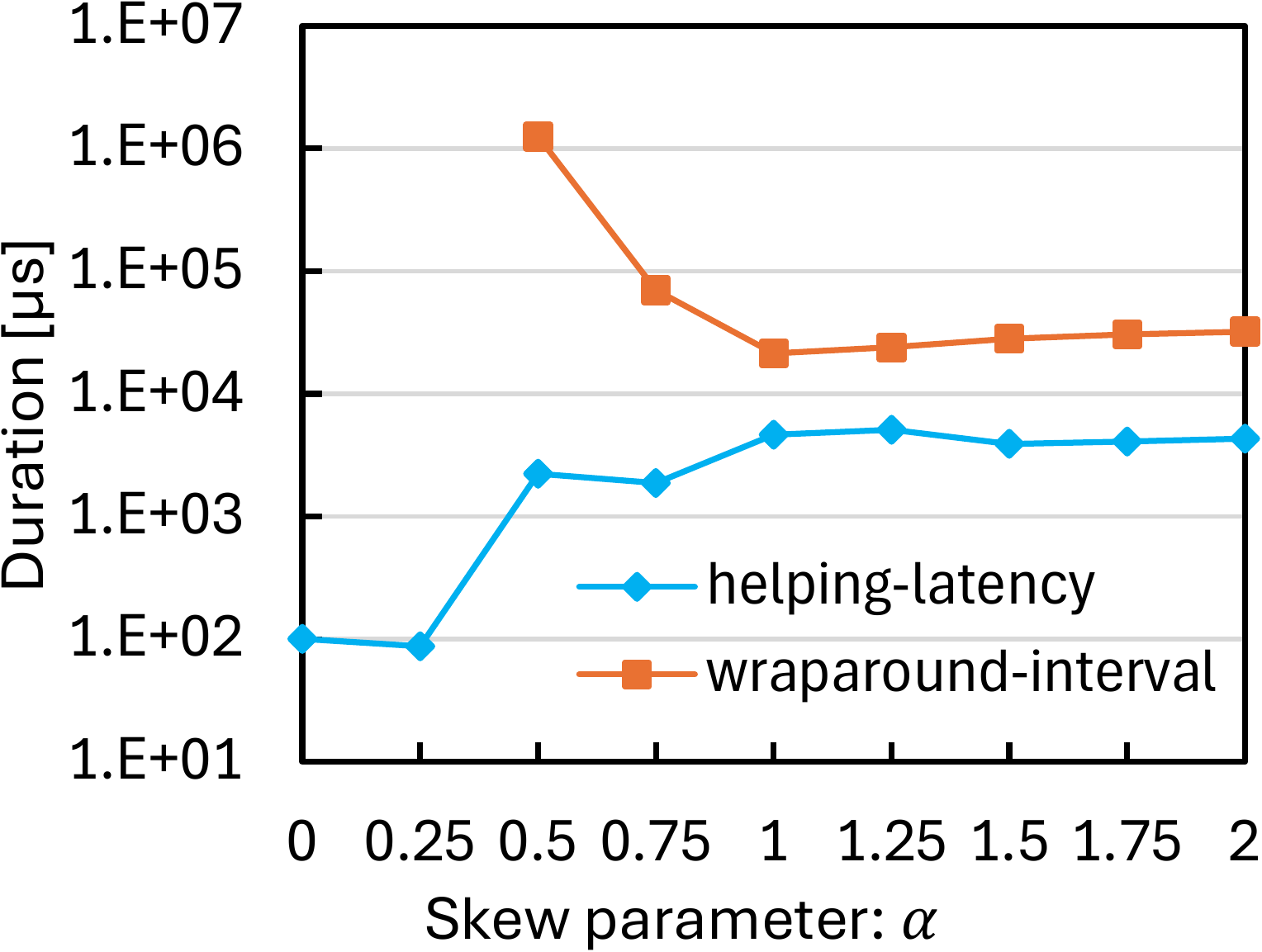}
        \subcaption{112 threads}
        \label{fig:skew_safety_112}
    \end{minipage}
    \hfill
    \begin{minipage}{0.495\linewidth}
        \centering
        \includegraphics[width=\linewidth]{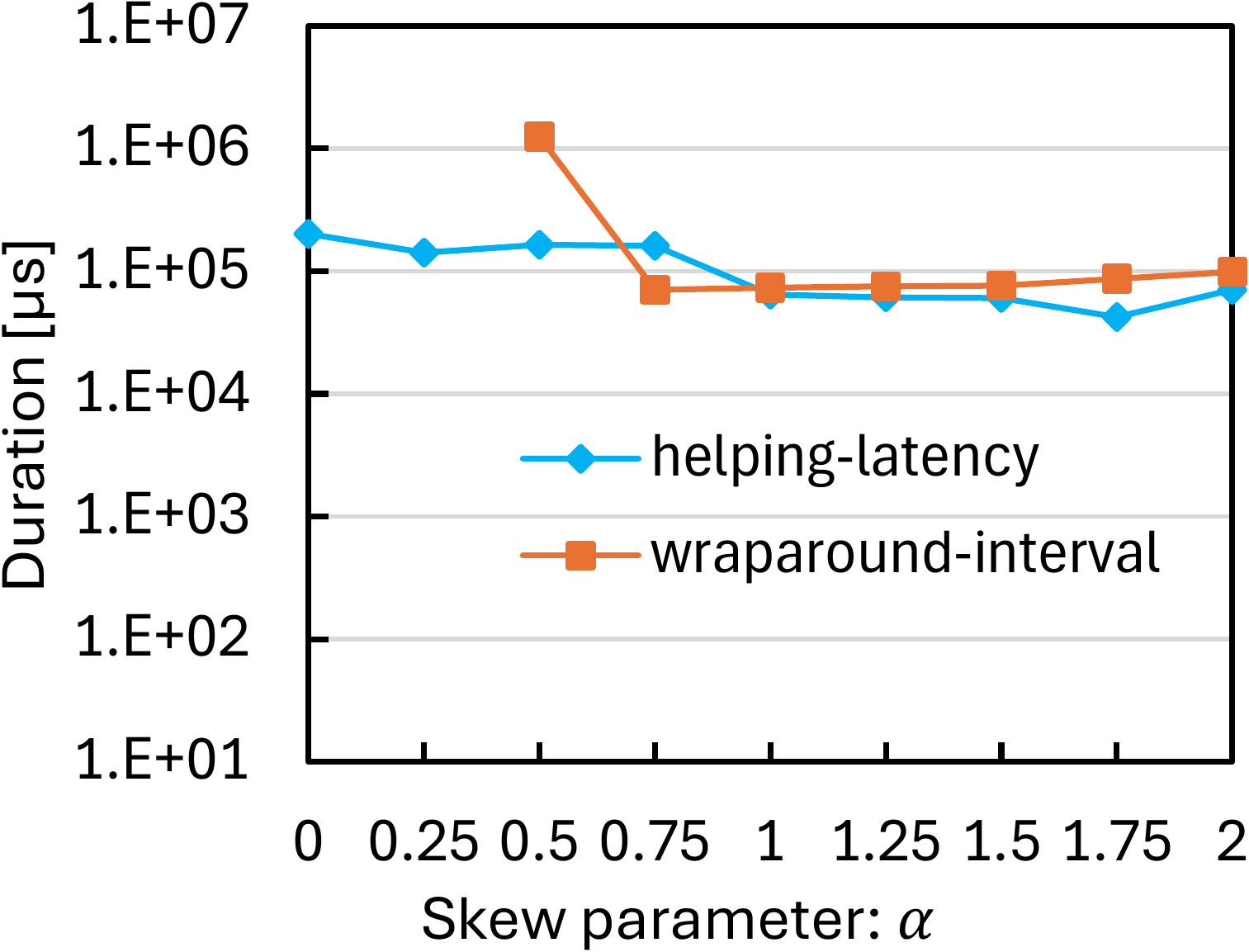}
        \subcaption{1024 threads}
        \label{fig:skew_safety_1024}
    \end{minipage}
    \caption{Measurements of helping latency and wraparound interval across different skew parameters.}
    \label{fig:skew_safety_combined}
\end{figure}

First, the changes in these metrics as the skew parameter varies are discussed.
\Cref{fig:skew_safety_combined} shows measurements with 112 and 1,024 threads.
These results show that the consistency is guaranteed across all skew parameters with 112 threads, but the ABA problem may occur with $\alpha = 0.75$ when using 1,024 threads.
With 1,024 threads, $\mathit{wraparound\_interval}$ drops sharply while $\mathit{helping\_latency}$ maintains large values regardless of skew parameters.
As a result, although $\mathit{helping\_latency}$ slightly decreases as the skew parameter increases, the two metrics invert at $\alpha = 0.75$.

\begin{figure}[t]
    \centering
    \begin{minipage}{0.495\linewidth}
        \centering
        \includegraphics[width=\linewidth]{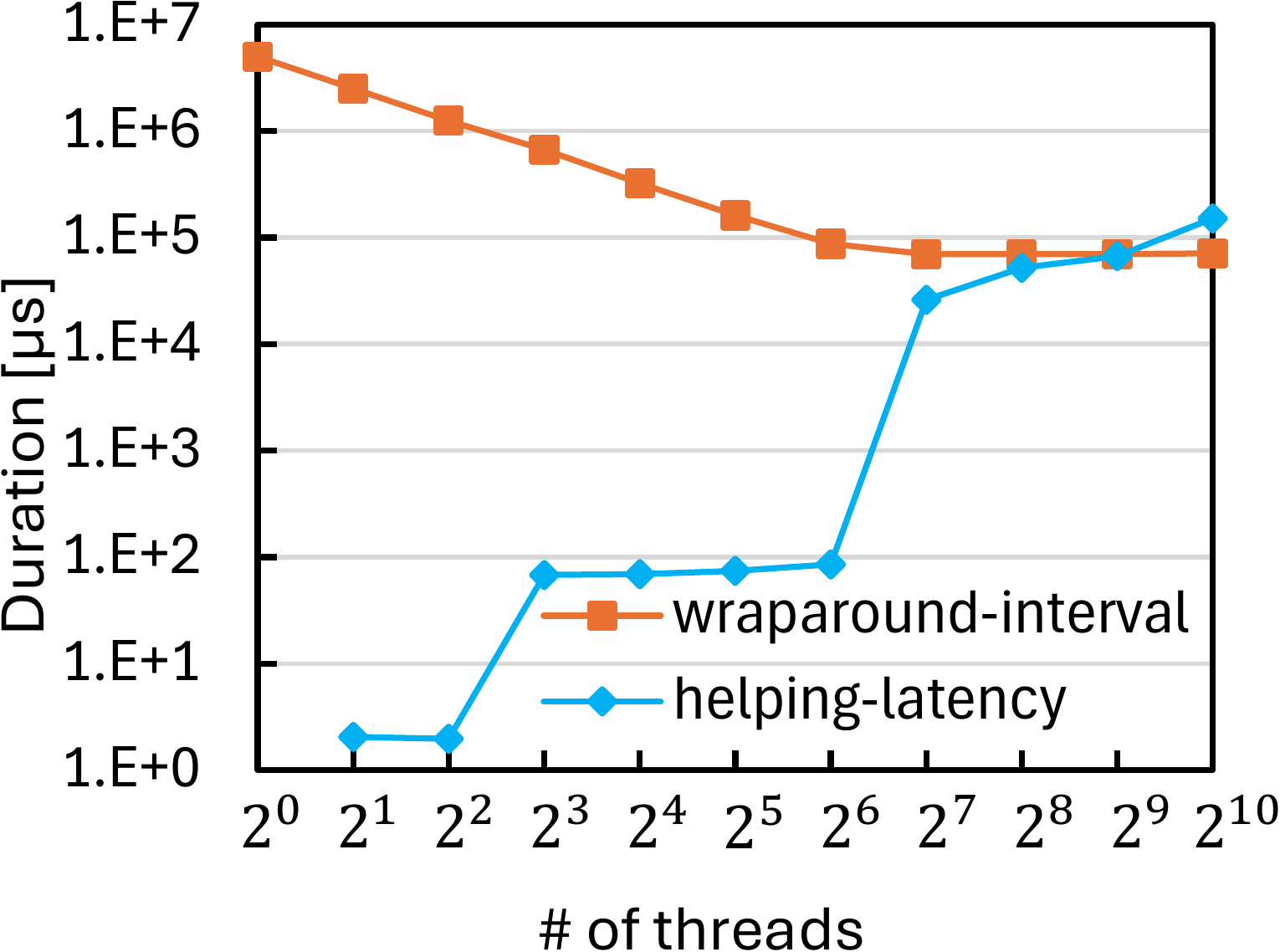}
        \subcaption{$\alpha=0.75$}
        \label{fig:thread_safety_075}
    \end{minipage}
    \hfill
    \begin{minipage}{0.495\linewidth}
        \centering
        \includegraphics[width=\linewidth]{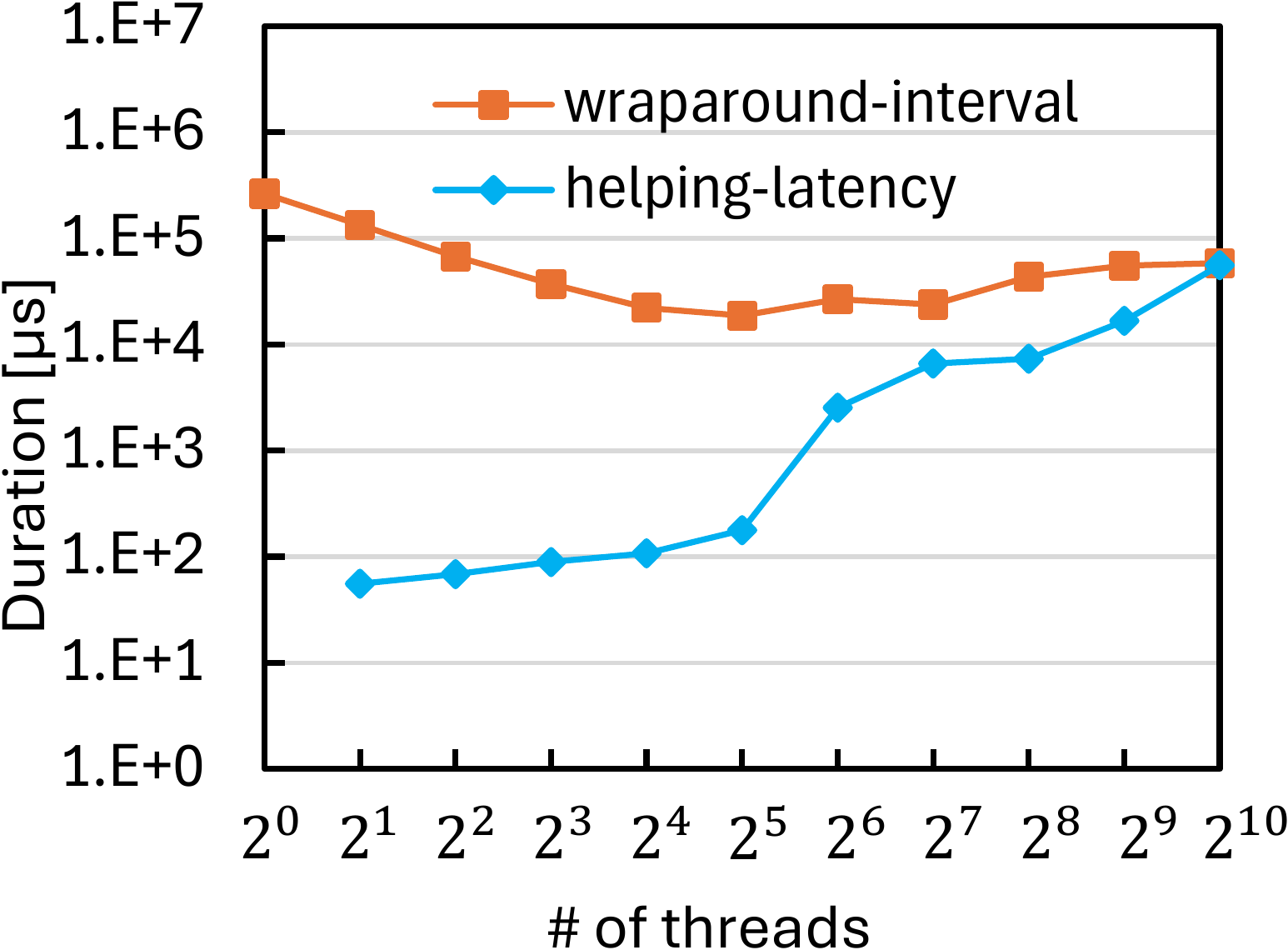}
        \subcaption{$\alpha=1.0$}
        \label{fig:thread_safety_100}
    \end{minipage}
    \caption{Measurements of helping latency and wraparound interval across different thread counts.}
    \label{fig:thread_safety_combined}
\end{figure}

However, since such excessive oversubscription is unrealistic, we also investigate metric changes across thread counts under specific skew parameters.
\Cref{fig:thread_safety_combined} shows results with $\alpha = 0.75$ and $\alpha = 1.0$.
As shown above, while the ABA problem may occur at 1,024 threads, the proposed method guarantees consistency at lower thread counts, even under oversubscription.
These results suggest that allocating 15 bits is reasonable for sufficient safety and practicality; the remaining bits are available for pointers to construct any data structures.
With respect to bit width, $\mathit{wraparound\_interval}$ scales exponentially, while $\mathit{helping\_latency}$ remains constant, ensuring greater safety predictability than existing lock-free algorithms.
Users can adjust the bit width based on their MCAS frequency.

\section{Conclusion}
\label{sec:conclusion}
In this paper, we propose a new lock-free MCAS algorithm.
The proposed method enhances efficiency by controlling the helping mechanism based on contention states.
Experimental results demonstrate that this contention-aware approach achieves up to three times the throughput of the state-of-the-art lock-free MCAS algorithm in high-contention environments.
Furthermore, this paper reveals logical flaws in existing MCAS algorithms that can lead to the ABA problem.
To address this vulnerability, we introduce version embedding and demonstrate its practicality through experiments.
However, version embedding does not completely eliminate the ABA problem.
Therefore, our future work includes exploring performance optimizations for lock-free MCAS while guaranteeing strict consistency.

\section*{Acknowledgments}
This work was partly supported by JSPS KAKENHI Grant Numbers JP23K24850, JP25K00161, JP25K21206, and JP26K02916.